\documentclass[aps,prl,twocolumn,superscriptaddress,showpacs]{revtex4}
\usepackage{longtable}
\usepackage{graphicx}% Include figure files
\usepackage{dcolumn}% Align table columns on decimal point
\usepackage{bm}% bold math
\usepackage{amsmath}
\usepackage[psamsfonts]{amssymb}

\newcommand{\STO}{SrTiO$_3$}
\newcommand{\LAO}{LaAlO$_3$}

\newcommand{\MR}{magnetoresistance}

\newcommand{\etal}{\emph{et al.}}

\begin{document}

\title{Magnetotransport effects in polar versus non-polar \STO~ based heterostructures}



\author{E. Flekser}

\affiliation{Raymond and Beverly Sackler School of Physics
and Astronomy, Tel-Aviv University, Tel Aviv, 69978, Israel}
\author{M. Ben Shalom}
\affiliation{Raymond and Beverly Sackler School of Physics
and Astronomy, Tel-Aviv University, Tel Aviv, 69978, Israel}
\author{M. Kim}
\affiliation{Stanford Institute for Materials and Energy Sciences, SLAC National Accelerator Laboratory, Menlo Park, CA 94025, USA}
\author{C. Bell}
\affiliation{Stanford Institute for Materials and Energy Sciences, SLAC National Accelerator Laboratory, Menlo Park, CA 94025, USA}
\author{Y. Hikita}
\affiliation{Stanford Institute for Materials and Energy Sciences, SLAC National Accelerator Laboratory, Menlo Park, CA 94025, USA}
\author{H. Y. Hwang}
\affiliation{Stanford Institute for Materials and Energy Sciences, SLAC National Accelerator Laboratory, Menlo Park, CA 94025, USA}
\author{Y. Dagan}
\affiliation{Raymond and Beverly Sackler School of Physics
and Astronomy, Tel-Aviv University, Tel Aviv, 69978, Israel}
\email[]{yodagan@post.tau.ac.il}


\date{\today}

\begin{abstract}
Anisotropic magnetoresistance and negative magnetoresistance for in-plane fields are compared for the \LAO/\STO~ interface and the symmetric Nb-doped \STO~ heterostructure. Both effects are exceptionally strong in \LAO/\STO. We analyze their temperature, magnetic field and gate voltage dependencies and find them to arise from a Rashba type spin-orbit coupling with magnetic scatterers that have two contributions to their potential: spin exchange and Coulomb interaction. Atomic spin-orbit coupling is sufficient to explain the small effects observed in Nb-doped \STO. These results clarify contradicting transport interpretations in \STO-based heterostructures.
\end{abstract}

%\keywords{}

%In nb/STO dresselhaus no magnetic scattering small effect persists up to high T no structural inversion symmetry breaking no gate dependence is related to band anisotropy.
% In STO/LAO magnetic+electric scattering therefore large effects. positive gate - Rashba + Dresselhous high T: dresselhous
%Negative gate -  Rashba only.

\maketitle
%\section {Introduction}
The interface between oxides can exhibit electronic properties that are  substantially different from those of its constituting materials\cite{OkamotoMillis}. The most studied example is the conducting interface formed between \STO~ and \LAO~
\cite{OhtomoHwang}, which exhibits superconductivity \cite{reyren2007superconducting} with a tunable critical temperature \cite{caviglia2008electric}. Various probes showed evidence for magnetic effects at the interface: hysteresis in the \MR~ at low temperatures \cite{brinkman2007magnetic, Dikin2011PhysRevLettCoexistence}, SQUID magnetometry \cite{wang2011electronic} showed ferromagnetic response for samples deposited at high oxygen pressure, and recently scanning SQUID microscopy \cite{bert2011direct} and torque magnetometry \cite{li2011coexistence} measurements suggested the coexistence of superconductivity and ferromagnetism. Theoretical model explaining this coexistence has been proposed \cite{michaeli2012superconducting}. It has been found that the magnetic state is not related to itinerant electrons but to localized moments that are independent of carrier concentration \cite{kalisky2012critical}. Magnetic moments were not observed in \STO/Nb-doped \STO/\STO~(Nb/\STO) heterostructures suggesting that magnetism is related to the polar discontinuity characterizing \LAO/\STO, or to a strong electric field applied at the surface \cite{LeePhysRevLett2011PhasediagSTO, lee2011electrolyte}. Another view based on studies of La-doped \STO~ system is that magnetism in \STO~ based systems appears at high carrier concentrations\cite{moetakef2012carrier}.
\par
\begin{figure}
\includegraphics[width=80mm]{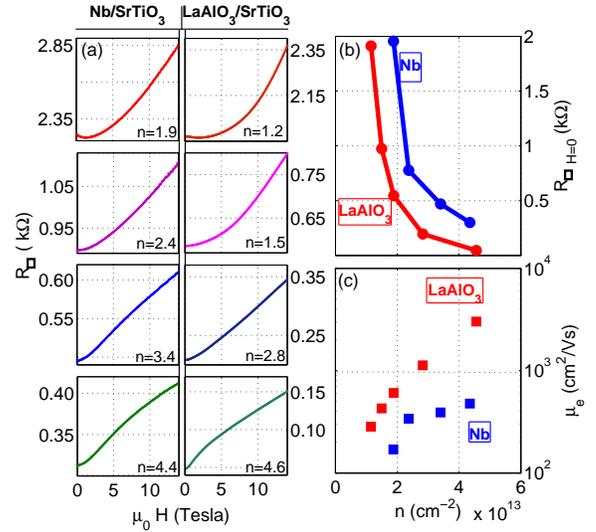}
\caption {Color on-line (a) Sheet resistance of \LAO/\STO~ (Sample 1) and Nb/\STO~ as a function of magnetic field for the various carrier concentrations n denoted in units of 10$^{13}$ cm$^{-2}$ and varied using back-gate voltage. (b) The sheet resistance at 2 K versus n for the two types of samples.(c) The mobility inferred from (b) as a function of n.\label{Transcomp}}
\end{figure}
While the superconducting state has a conspicuous signature in transport measurements, such a signature for the magnetic state remains illusive. M. Ben Shalom \etal~ showed that a strong \MR~ anisotropy is observed together with superconductivity. They interpreted their data in terms of anisotropic spin scattering resulting from magnetic ordering together with a strong spin-orbit interaction\cite{Sachs2010S746, shalom2009anisotropic}. This spin-orbit interaction was measured using its effect on the transport properties in the normal and in the superconducting state \cite{ben2010tuning} and in the weak localization regime \cite{caviglia2010tunable}. Furthermore, Seri and Klein suggested that the anomalous behavior of the Hall signal is a result of a field induced magnetism \cite{seri:180410}.
\par
There are various \STO~ based systems that exhibit properties of a two dimensional electron gas \cite{lee2011electrolyte, santander2011two, meevasana2011creation, xie2011tuning, ngai2010electric, biscaras2010two, ye2009liquid, LeePhysRevLett2011PhasediagSTO}. Of these systems Nb/\STO~ has the most symmetric structure, it also lacks polarization existing in the other systems. It is therefore worth to carefully compare the anisotropy of the transport properties of Nb/\STO~ and \LAO/\STO~.
\par
Superconductivity and spin-orbit interaction in Nb/\STO~ were studied by M. Kim \etal~\cite{kim2011intrinsic}. They found that the upper critical field exceeds the Clogston-Chandrasekhar limit with values similar to \LAO/\STO. They suggested that the spin-orbit interaction plays an important role also in Nb/\STO, thus raising questions on a purely Rashba type spin-orbit coupling model for the various systems.
\par
In the study we present here, we compare the \MR~ and anisotropic \MR~ (AMR) in the two dimensional electron gas in heterostructures of Nb/\STO~ and \LAO/\STO. The large negative \MR~observed for in-plane magnetic fields (nPMR) in \LAO/\STO~ is absent in Nb/\STO~ for the carrier densities under study. We show that the AMR observed when rotating the in-plane magnetic field with respect to the current direction and the nPMR arise from the same mechanism. From the sign and magnitude of the AMR effect we conclude that polarized magnetic scatterers and a Rashba type spin-orbit interaction are responsible for this effect. This scattering process is absent in the symmetric structure of Nb/\STO~ where a smaller and opposite in sign effect is seen.
\par
%Besides STO/LAO interface many other STO based systems exhibit unusual transport properties: Ionic liquid gated single STO %crystals \cite{lee2011electrolyte}, vacuum cleaved STO \cite{santander2011two}, UV irradiated STO surface %\cite{meevasana2011creation}, Nb/STO \cite{xie2011tuning}, Ar irradiated STO surfaces \cite{ngai2010electric} and KTiO$_3$/STO %\cite{biscaras2010two}. Of these STO based systems Nb/STO has the most symmetric structure, it also lacks intrinsic polarization %existing in the interfaces. It is therefore worth to carefully compare the anisotropy of the transport properties of Nb/STO and %LAO/STO. This will allow us to differentiate between effects that arise from the asymmetric polar structure.
\par

%We measured both materials, out of plane field and gate dependence transport properties are similar.
%In plane field properties are different: Negative parallel and AMR.
%Magnetization\Rashba in sto/lao not in nb doped: negative versus positive AMR versus null
%Can we scale the gate dependence for various samples?
%(Look at Hso from 2DWL).
%
\par %sample description
Thin films of \LAO~ were deposited by a pulsed laser on atomically flat \STO~ substrates with : 6, 8 and 10 unit cells respectively. The oxygen pressure during the deposition was maintained at $10^{-4}$ Torr and the temperature was 785 $^\circ$C. The deposition was
followed by a two-hour annealing stage at oxygen pressure of 0.2
Torr and a temperature of 500$^\circ$ C. The thickness of the \LAO~layer was monitored by
reflection high-energy electron diffraction. The samples were patterned into narrow strips or into Hall bars \cite{schneiderlithography}. A Nb/\STO~ sample with 1\% Nb 6.7 nm thick with 100 nm \STO~ buffer and capping layers was deposited as described elsewhere \cite{kozuka2009two}. A layer of
gold was evaporated at the bottom of the sample and used as a gate
when biased to $\pm$100 V relative to the 2DEG. Contacts to the \LAO/\STO~ samples were made
by drilling holes through the \LAO~ layer using ion irradiation followed by sputtering Ti/Al or Ti/Au pads.
\par
We have previously showed that in the presence of a large parallel magnetic field a small perpendicular component can result in a dramatic increase in resistance \cite{shalom2009anisotropic}. A small deviation from exact parallel position (e.g. due to a wobble in the rotator probe) can therefore have a strong undesired effect on the data. In order to avoid this spurious effect we simultaneously measured two perpendicular bridges on the same substrate and made sure that all reported effects are phase shifted by 90$^\circ$ between the two bridges. For example for current running along the [110] direction we had another bridge with current running along the [1$\bar{1}$0] direction.
\par
%Figure 1
In Fig.\ref{Transcomp}a we compare the sheet resistance versus magnetic field curves for \LAO/\STO~ and Nb/\STO~ at 2 K. Here the magnetic field is applied perpendicular to the conducting plane. The carrier concentration is varied by the gate voltage and estimated from the high field slope of the Hall resistance ($\mu_0 H >$12 T). For \LAO/\STO~ the typical dependence of longitudinal and Hall resistivity on gate and temperature is observed \cite{bell2009dominant}. The dependencies of the sheet resistance and mobility as a function of carrier concentration are shown in Fig.\ref{Transcomp}b.
%Figure2
\begin{figure}
\includegraphics[width=80mm]{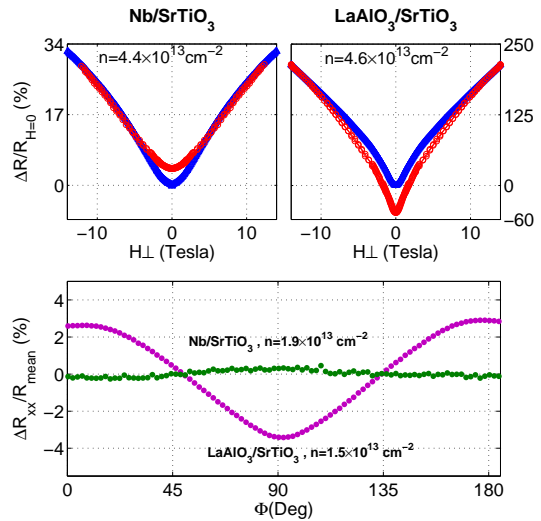}
\caption {(Color on-line) Upper panel: Magnetoresistance comparison between magnetic field applied perpendicular to the plane swept between -14 to 14 T (blue triangles) and rotation in a constant field of 14 T thus varying the perpendicular component (red circles) for Nb/\STO~ and \LAO/\STO~ (Sample 1) at 2 K. Negative \MR~ is observed only for \LAO/\STO~. Lower panel: Anisotropic magnetoresistance for in-plane field of 18 T rotated with respect to the current direction at 20 mK. Large positive AMR is observed for \LAO/\STO~. A smaller effect is observed for Nb/\STO~ with an opposite sign.\label{fig2}}
\end{figure}
The first significant difference between Nb/\STO~ and \LAO/\STO~ is revealed when applying a magnetic field component parallel to the plane. In Fig.2 (upper panel) we show two data sets for each sample: triangles are data taken while sweeping the magnetic field perpendicular to the conducting plane. The circles represent data taken while holding a constant magnetic field of 14 T and changing the angle $\theta$ between the field and the plane. The sheet resistance is plotted against the perpendicular field amplitude (component) for the field sweep (rotation) experiments. While for the \LAO/\STO~ a strong pronounced negative \MR~is seen when a significant parallel field component is present, such a component contributes a merely small positive \MR~ for the Nb/\STO~ sample. For example, $H_{\perp}$=0 for the circles corresponds to 14 T applied parallel to the plane resulting in a positive \MR~ for the Nb/\STO~ and a strong negative \MR~ for \LAO/\STO. The small in-plane positive \MR~ in Nb/\STO~ is a result of the 90$^\circ$ angle between the in-plane field and the current. The absence of nPMR in Nb/\STO~ holds for all carrier concentrations studied here and is one of the key findings reported in this Letter.
\begin{figure}
\includegraphics[width=80mm]{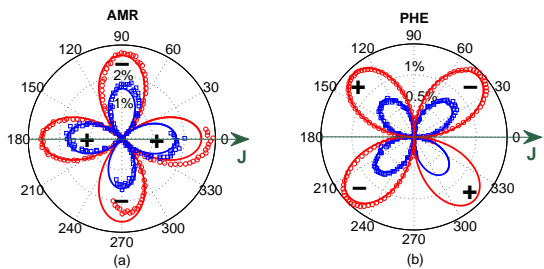}
\caption {Color on-line (a) The anisotropic magnetoresistance normalized to its mean value (Sample 1), $\frac{\Delta R_\textrm{{xx}}}{R_\textrm{{mean}}}$ as a function of $\phi$ the angle between the current and the magnetic field (18 T) at 20 mK in polar representation. Data are presented for two carrier concentrations (controlled by the gate voltage) circles $n=1.8\times10^{13}$ cm$^{-2}$ blue squares $n=1.3\times10^{13}$ cm$^{-2}$ the solid lines are fits to $\cos^2(\phi)$ (b) The transverse voltage signal normalized to the mean longitudinal voltage ($V_{xx}$) in (a). This planar Hall signal is shifted by 45$^\circ$ relative to the longitudinal signal as expected. Solid lines are fits to $\cos^2(\phi-\frac{\pi}{4})$.\label{fig3}}
\end{figure}
\par
Rotating the in-plane magnetic field with respect to the current unravels another important difference between Nb/\STO~ and \LAO/\STO. The AMR amplitude, $\Delta R_{AMR}=\frac{\rho_\parallel-\rho_\bot}{(\rho_\parallel+\rho_\bot)/2}$, is large and positive for \LAO/\STO~ (i.e. maximum resistance for $\vec{H}\|\vec{J}$ and minimum for $\vec{H}\bot\vec{J}$ with $\vec{J}$ being the current density). By contrast, for Nb/\STO~ it is very small and negative.
%Since the positive large AMR effect exists only in STO/LAO we now describe its dependence on gate, field, temperature and current orientation with respect to the crystal axis.
\par
\begin{figure}
\includegraphics[width=80mm]{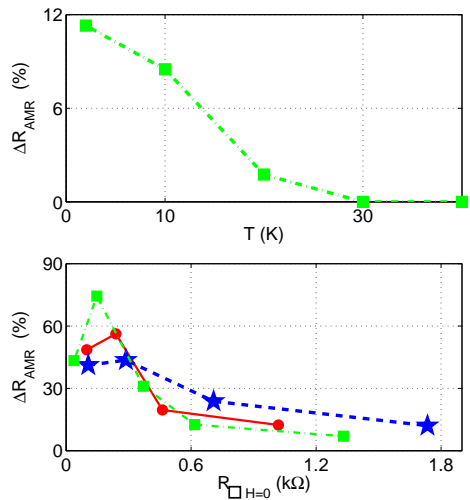}
\caption {Color on-line. Upper panel AMR amplitude: $\Delta R_{AMR}=\frac{\rho_\parallel-\rho_\bot}{(\rho_\parallel+\rho_\bot)/2}$ as a function of temperature. Lower panel: AMR amplitude as a function of the zero field resistance controlled by gate voltage for various samples: Green squares (Sample 1) are data taken at 20 mK and 18 T for current along [100]. Red circles (Sample 2) and blue stars (Sample 3) are data taken at 2 K and 14 T for bridges along [110] (red) and [100] (blue). (see also supplementary part).\label{fig4}}
\end{figure}
Anisotropy in the longitudinal conductivity (such as the AMR) should result in a planar Hall signal. In this configuration the transverse voltage signal is measured for in-plane fields as a function of the angle between the field and current. This signal should be phase shifted by $45^{\circ}$ compared to the longitudinal AMR \cite{tang2003giant}.
\par
In Fig.3a we show the AMR for two gate voltages in polar representation. Note that the angular dependence follows $\cos^2\phi$ with $\phi$ the angle between the current and the field. The corresponding planar Hall effect (PHE) is shown in Fig.3b. The fact that the AMR and PHE signal are observed with the proper phase shift of 45$^\circ$ and with an amplitude ratio of $1/2$ ensures that the anisotropy is an intrinsic property, unrelated to a spurious measurement artifact.
\par
In Fig.4a we study the temperature dependence of the AMR. The AMR disappears at $\sim$30 K for all gate voltages. This is the temperature above which the nPMR disappears, consistent with our previous observations \cite{shalom2009anisotropic}. The fact that both AMR and the nPMR vanish at the same temperature suggests that they are two expressions of the same scattering mechanism.
%The next step is to study the evolution of both effects with carrier concentration. This is done in Fig.4b.
\par
The amplitude of the AMR is shown versus the sheet resistance in Fig.4b. for three different samples in the maximal accessible field. Here we use the sheet resistance as a useful parameter (monotonic with carrier concentration) for comparison of different samples. We note that no significant difference in the AMR amplitude is observed when the current direction is along [110] or [100] (See supplementary information for more details). This suggests that the most important parameter affecting the AMR is the angle between the current and field. We emphasize that standard defects cannot result in such strong variation in AMR amplitude with gate: from a few percent at low carrier concentration to 85\%.
\par
\begin{figure}
\includegraphics[width=80mm]{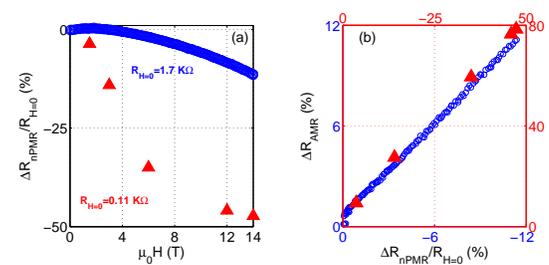}
\caption {Color on-line (a) nPMR effect $\frac{\Delta R_{xx}(H_\|)}{R_{0}}$ as a function of magnetic field (Sample 3) for two gate voltages at 2 K. (b) AMR versus the nPMR for the same gate voltages. We note that the two effects scale.\label{fig5}}
\end{figure}
The final evidence that the nPMR and AMR originate from the same scattering mechanism is found from their field dependence as demonstrated in Fig.5. We show two data sets of high and low carrier concentrations for the same sample. The nPMR exhibits very different field dependence (Left hand panel) for these two concentrations. Furthermore, for the high carrier concentration the field dependence of the nPMR  has a few characteristic features: starting with a flat region followed by a steep descent and a saturation regime \cite{ben2010tuning}. Nevertheless, the amplitude of the AMR tracks this behavior. The two effects have similar field dependence up to a numerical coefficient. This is clear from Fig.5b. where we plot the amplitude of the AMR versus the nPMR for the above gate voltages. Indeed a linear dependence is obtained.
\par
Since both nPMR and AMR effects exhibit the same temperature and magnetic field dependencies for various carrier concentrations they must arise from the same scattering mechanism. This key observation allows us to focus on the mechanism responsible for the AMR and thus the other magnetotransport effects.
\par
Three possible mechanisms for the AMR have been proposed \cite{Vybomy2009MicroscopicmechanismAMR}: a. anisotropic Fermi velocities coupled to a finite magnetization through the spin-orbit interaction, b. spin-orbit coupled magnetically polarized carriers with isotropic scattering centers, and c. anisotropic magnetic scattering centers acting on unpolarized spin-orbit coupled carriers. Mechanisms a. and b. usually result in a smaller AMR effect since the magnetization and the spin-orbit interaction compete with each other. This is not the case in mechanism c. where the magnetization and the spin orbit interaction do not exist in the same band states, thus allowing fully polarized scatterers despite the strong spin-orbit interaction. The unusually large AMR (up to 85\%) suggests that the dominant mechanism in \LAO/\STO~ is mechanism c.
\par
The nature of the scattering centers in mechanism c. can be either purely magnetic (only spin exchange) or electro-magnetic (combined with the Coulomb interaction). Large AMR can be observed only if the Fermi energy, E$_F$ is much smaller than the spin-orbit interaction for the first case. Calculations for the case of electro-magnetic scatterers show that the AMR amplitude can be very large (twice as large as for the pure magnetic case) even for a spin-orbit interaction smaller than E$_F$. In this case the amplitude of the AMR strongly depends on the ratio between the electric and the magnetic parts of the scattering potential, with a sharp maximum at the point where this ratio becomes unity. The origin of this large AMR is a destructive interference between the magnetic and the non-magnetic back-scattering amplitudes \cite{Trushin2009AMRpolaraizedspins}.
\par
As can be seen in Fig.4b the AMR is large for a broad range of concentrations and sheet resistances. Since in \LAO/\STO~ the spin-orbit interaction is of the order of E$_F$, the scattering potential should include an electric part. This suggests that localized charges act as both electric and magnetic scattering centers and the maximum AMR should not necessarily coincide with the maximum of spin-orbit interaction. Upon changing the gate voltage we identify the maximum in AMR as the point where the magnetic and electric scattering amplitude ratio becomes of the order one. Moving away from this point will alter this ratio by either screening the electric scattering or by moving the wave function away from the interface and consequently reduce the coupling between the magnetic scatterers and the conduction electrons.
\par
A more detailed description of our system should include the way the spin-orbit interaction changes with carrier concentration. In addition,as the carrier density increases the $d_{yz}$ and $d_{xz}$ bands become populated and the anisotropy of the band mass may become important \cite{joshua2011universal}.
\par
Spin-orbit interaction at the atomic level without coupling the band momentum to the spin is not expected to result in AMR. Large AMR amplitude is expected if the Rashba or the Dresselhaus effects take place. However, the sign of the AMR is positive, i.e. maximum resistance for magnetic field parallel to the current for the Rashba and negative for the Dresselhous one \cite{Trushin2009AMRpolaraizedspins}. This allows us to conclude that the main contribution to the AMR in \LAO/\STO~ comes from the Rashba effect.
\par
For Nb/\STO~ large spin-orbit coupling has been suggested based on the large $H_{c2}$ parallel magnetic field needed to quench superconductivity \cite{kim2011intrinsic}. However, in this case the AMR is very small and negative. We therefore explain the large $H_{c2}$ observed by M. Kim \etal~ as a result of atomic spin-orbit interaction\cite{SpinorbitHcpar}. The small negative AMR can be a result of a finite thickness of the confinement zone \cite{shalom2009anisotropic}. For Nb/\STO~ at the carrier concentration studied, both magnetism and the Rashba term do not play a significant role in transport.
\par
A negative contribution to the AMR can be seen in \LAO/\STO~at low temperatures for large carrier concentrations.\cite{shalom2009anisotropic} In these concentrations the AMR deviates from the simple $\cos^2(\phi)$ and becomes sharp. This effect is seen both for current running along [100] and [110] directions. It is therefore less likely to be related to a crystalline magnetic anisotropy.
\par
In summary, we show that anisotropic \MR~ (AMR) and planar Hall effect are observed in \LAO/\STO~ when rotating the magnetic field in-plane with respect to the current. We show that the AMR scales with the in-plane negative \MR~(nPMR). This suggests that both effects arise from the same spin scattering mechanism combined with a strong spin-orbit interaction. Importantly, the AMR is very small and negative and the nPMR is absent in the non-polar symmetric structure of Nb-doped \STO~ for the carrier range under study. Since the AMR has similar amplitudes for current flowing along different crystallographic directions: $\vec{J}\|$[110] or $\vec{J}\|$[100], the important parameter governing it is the angle between field and current. The sign of the effect in \LAO/\STO~ shows that the spin-orbit interaction involved in the scattering process is of the Rashba type. This is not the case in Nb/\STO~ where atomic spin-orbit interaction is sufficient to explain the small effects observed. The large magnitude of the AMR effect in the polar structure is explained by a scattering potential with two components: spin exchange and Coulomb interaction. The ratio between these contributions varies with gate voltage, which results in a strong gate dependence. The fact that mobile carriers and the magnetic moments do not occupy the same band state may allow coexistence of superconductivity and magnetism.
\par
We thank K. Michaeli, K. Vyborny, S. Ilani and I. Neder for useful discussions. This research was partially supported by Grant No. 2010140 from the United States-Israel Binational Science Foundation; by the Infrastructure program of the Israeli Ministry of Science and Technology and by the Israel Science Foundation under grant 1421/08. M.K., C.B. Y.H. and H.Y.H. acknowledge support by the Department of Energy, Office of Basic Energy Sciences, Materials Sciences and Engineering Division, under contract DE-AC02-76SF00515. A portion of this work was performed at the National High Magnetic Field Laboratory, which is supported by NSF Cooperative Agreement No. DMR-0654118, by the State of Florida, and by the DOE.

\bibliographystyle{apsrev}

\bibliography{spinorbit9}

\begin{thebibliography}{34}
\expandafter\ifx\csname natexlab\endcsname\relax\def\natexlab#1{#1}\fi
\expandafter\ifx\csname bibnamefont\endcsname\relax
  \def\bibnamefont#1{#1}\fi
\expandafter\ifx\csname bibfnamefont\endcsname\relax
  \def\bibfnamefont#1{#1}\fi
\expandafter\ifx\csname citenamefont\endcsname\relax
  \def\citenamefont#1{#1}\fi
\expandafter\ifx\csname url\endcsname\relax
  \def\url#1{\texttt{#1}}\fi
\expandafter\ifx\csname urlprefix\endcsname\relax\def\urlprefix{URL }\fi
\providecommand{\bibinfo}[2]{#2}
\providecommand{\eprint}[2][]{\url{#2}}

\bibitem[{\citenamefont{Okamoto and Millis}(2004)}]{OkamotoMillis}
\bibinfo{author}{\bibfnamefont{S.}~\bibnamefont{Okamoto}} \bibnamefont{and}
  \bibinfo{author}{\bibfnamefont{A.~J.} \bibnamefont{Millis}},
  \bibinfo{journal}{Nature (London)} \textbf{\bibinfo{volume}{428}},
  \bibinfo{pages}{630} (\bibinfo{year}{2004}).

\bibitem[{\citenamefont{Ohtomo and Hwang}(2004)}]{OhtomoHwang}
\bibinfo{author}{\bibfnamefont{A.}~\bibnamefont{Ohtomo}} \bibnamefont{and}
  \bibinfo{author}{\bibfnamefont{H.~Y.} \bibnamefont{Hwang}},
  \bibinfo{journal}{Nature (London)} \textbf{\bibinfo{volume}{427}},
  \bibinfo{pages}{423} (\bibinfo{year}{2004}).

\bibitem[{\citenamefont{Reyren et~al.}(2007)\citenamefont{Reyren, Thiel,
  Caviglia, Kourkoutis, Hammerl, Richter, Schneider, Kopp, R{\"u}etschi,
  Jaccard et~al.}}]{reyren2007superconducting}
\bibinfo{author}{\bibfnamefont{N.}~\bibnamefont{Reyren}},
  \bibinfo{author}{\bibfnamefont{S.}~\bibnamefont{Thiel}},
  \bibinfo{author}{\bibfnamefont{A.}~\bibnamefont{Caviglia}},
  \bibinfo{author}{\bibfnamefont{L.}~\bibnamefont{Kourkoutis}},
  \bibinfo{author}{\bibfnamefont{G.}~\bibnamefont{Hammerl}},
  \bibinfo{author}{\bibfnamefont{C.}~\bibnamefont{Richter}},
  \bibinfo{author}{\bibfnamefont{C.}~\bibnamefont{Schneider}},
  \bibinfo{author}{\bibfnamefont{T.}~\bibnamefont{Kopp}},
  \bibinfo{author}{\bibfnamefont{A.}~\bibnamefont{R{\"u}etschi}},
  \bibinfo{author}{\bibfnamefont{D.}~\bibnamefont{Jaccard}},
  \bibnamefont{et~al.}, \bibinfo{journal}{Science}
  \textbf{\bibinfo{volume}{317}}, \bibinfo{pages}{1196} (\bibinfo{year}{2007}).

\bibitem[{\citenamefont{Caviglia et~al.}(2008)\citenamefont{Caviglia, Gariglio,
  Reyren, Jaccard, Schneider, Gabay, Thiel, Hammerl, Mannhart, and
  Triscone}}]{caviglia2008electric}
\bibinfo{author}{\bibfnamefont{A.}~\bibnamefont{Caviglia}},
  \bibinfo{author}{\bibfnamefont{S.}~\bibnamefont{Gariglio}},
  \bibinfo{author}{\bibfnamefont{N.}~\bibnamefont{Reyren}},
  \bibinfo{author}{\bibfnamefont{D.}~\bibnamefont{Jaccard}},
  \bibinfo{author}{\bibfnamefont{T.}~\bibnamefont{Schneider}},
  \bibinfo{author}{\bibfnamefont{M.}~\bibnamefont{Gabay}},
  \bibinfo{author}{\bibfnamefont{S.}~\bibnamefont{Thiel}},
  \bibinfo{author}{\bibfnamefont{G.}~\bibnamefont{Hammerl}},
  \bibinfo{author}{\bibfnamefont{J.}~\bibnamefont{Mannhart}}, \bibnamefont{and}
  \bibinfo{author}{\bibfnamefont{J.}~\bibnamefont{Triscone}},
  \bibinfo{journal}{Nature} \textbf{\bibinfo{volume}{456}},
  \bibinfo{pages}{624} (\bibinfo{year}{2008}).

\bibitem[{\citenamefont{Brinkman et~al.}(2007)\citenamefont{Brinkman, Huijben,
  Van~Zalk, Huijben, Zeitler, Maan, Van~der Wiel, Rijnders, Blank, and
  Hilgenkamp}}]{brinkman2007magnetic}
\bibinfo{author}{\bibfnamefont{A.}~\bibnamefont{Brinkman}},
  \bibinfo{author}{\bibfnamefont{M.}~\bibnamefont{Huijben}},
  \bibinfo{author}{\bibfnamefont{M.}~\bibnamefont{Van~Zalk}},
  \bibinfo{author}{\bibfnamefont{J.}~\bibnamefont{Huijben}},
  \bibinfo{author}{\bibfnamefont{U.}~\bibnamefont{Zeitler}},
  \bibinfo{author}{\bibfnamefont{J.}~\bibnamefont{Maan}},
  \bibinfo{author}{\bibfnamefont{W.}~\bibnamefont{Van~der Wiel}},
  \bibinfo{author}{\bibfnamefont{G.}~\bibnamefont{Rijnders}},
  \bibinfo{author}{\bibfnamefont{D.}~\bibnamefont{Blank}}, \bibnamefont{and}
  \bibinfo{author}{\bibfnamefont{H.}~\bibnamefont{Hilgenkamp}},
  \bibinfo{journal}{Nature Materials} \textbf{\bibinfo{volume}{6}},
  \bibinfo{pages}{493} (\bibinfo{year}{2007}).

\bibitem[{\citenamefont{Dikin et~al.}(2011)\citenamefont{Dikin, Mehta, Bark,
  Folkman, Eom, and Chandrasekhar}}]{Dikin2011PhysRevLettCoexistence}
\bibinfo{author}{\bibfnamefont{D.~A.} \bibnamefont{Dikin}},
  \bibinfo{author}{\bibfnamefont{M.}~\bibnamefont{Mehta}},
  \bibinfo{author}{\bibfnamefont{C.~W.} \bibnamefont{Bark}},
  \bibinfo{author}{\bibfnamefont{C.~M.} \bibnamefont{Folkman}},
  \bibinfo{author}{\bibfnamefont{C.~B.} \bibnamefont{Eom}}, \bibnamefont{and}
  \bibinfo{author}{\bibfnamefont{V.}~\bibnamefont{Chandrasekhar}},
  \bibinfo{journal}{Phys. Rev. Lett.} \textbf{\bibinfo{volume}{107}},
  \bibinfo{pages}{056802} (\bibinfo{year}{2011}).

\bibitem[{\citenamefont{Wang et~al.}(2011)\citenamefont{Wang, Baskaran, Liu,
  Huijben, Yi, Annadi, Barman, Rusydi, Dhar, Feng et~al.}}]{wang2011electronic}
\bibinfo{author}{\bibfnamefont{X.}~\bibnamefont{Wang}},
  \bibinfo{author}{\bibfnamefont{G.}~\bibnamefont{Baskaran}},
  \bibinfo{author}{\bibfnamefont{Z.}~\bibnamefont{Liu}},
  \bibinfo{author}{\bibfnamefont{J.}~\bibnamefont{Huijben}},
  \bibinfo{author}{\bibfnamefont{J.}~\bibnamefont{Yi}},
  \bibinfo{author}{\bibfnamefont{A.}~\bibnamefont{Annadi}},
  \bibinfo{author}{\bibfnamefont{A.}~\bibnamefont{Barman}},
  \bibinfo{author}{\bibfnamefont{A.}~\bibnamefont{Rusydi}},
  \bibinfo{author}{\bibfnamefont{S.}~\bibnamefont{Dhar}},
  \bibinfo{author}{\bibfnamefont{Y.}~\bibnamefont{Feng}}, \bibnamefont{et~al.},
  \bibinfo{journal}{Nature Communications} \textbf{\bibinfo{volume}{2}},
  \bibinfo{pages}{188} (\bibinfo{year}{2011}).

\bibitem[{\citenamefont{Bert et~al.}(2011)\citenamefont{Bert, Kalisky, Bell,
  Kim, Hikita, Hwang, and Moler}}]{bert2011direct}
\bibinfo{author}{\bibfnamefont{J.~A.} \bibnamefont{Bert}},
  \bibinfo{author}{\bibfnamefont{B.}~\bibnamefont{Kalisky}},
  \bibinfo{author}{\bibfnamefont{C.}~\bibnamefont{Bell}},
  \bibinfo{author}{\bibfnamefont{M.}~\bibnamefont{Kim}},
  \bibinfo{author}{\bibfnamefont{Y.}~\bibnamefont{Hikita}},
  \bibinfo{author}{\bibfnamefont{H.~Y.} \bibnamefont{Hwang}}, \bibnamefont{and}
  \bibinfo{author}{\bibfnamefont{K.~A.} \bibnamefont{Moler}},
  \bibinfo{journal}{Nature Physics} \textbf{\bibinfo{volume}{7}},
  \bibinfo{pages}{767} (\bibinfo{year}{2011}).

\bibitem[{\citenamefont{Li et~al.}(2011)\citenamefont{Li, Richter, Mannhart,
  and Ashoori}}]{li2011coexistence}
\bibinfo{author}{\bibfnamefont{L.}~\bibnamefont{Li}},
  \bibinfo{author}{\bibfnamefont{C.}~\bibnamefont{Richter}},
  \bibinfo{author}{\bibfnamefont{J.}~\bibnamefont{Mannhart}}, \bibnamefont{and}
  \bibinfo{author}{\bibfnamefont{R.}~\bibnamefont{Ashoori}},
  \bibinfo{journal}{Nature Physics} \textbf{\bibinfo{volume}{7}},
  \bibinfo{pages}{762} (\bibinfo{year}{2011}).

\bibitem[{\citenamefont{Michaeli et~al.}(2012)\citenamefont{Michaeli, Potter,
  and Lee}}]{michaeli2012superconducting}
\bibinfo{author}{\bibfnamefont{K.}~\bibnamefont{Michaeli}},
  \bibinfo{author}{\bibfnamefont{A.}~\bibnamefont{Potter}}, \bibnamefont{and}
  \bibinfo{author}{\bibfnamefont{P.}~\bibnamefont{Lee}},
  \bibinfo{journal}{Physical Review Letters} \textbf{\bibinfo{volume}{108}},
  \bibinfo{pages}{117003} (\bibinfo{year}{2012}).

\bibitem[{\citenamefont{Kalisky et~al.}(2012)\citenamefont{Kalisky, Bert,
  Klopfer, Bell, Sato, Hosoda, Hikita, Hwang, and Moler}}]{kalisky2012critical}
\bibinfo{author}{\bibfnamefont{B.}~\bibnamefont{Kalisky}},
  \bibinfo{author}{\bibfnamefont{J.~A.} \bibnamefont{Bert}},
  \bibinfo{author}{\bibfnamefont{B.~B.} \bibnamefont{Klopfer}},
  \bibinfo{author}{\bibfnamefont{C.}~\bibnamefont{Bell}},
  \bibinfo{author}{\bibfnamefont{H.~K.} \bibnamefont{Sato}},
  \bibinfo{author}{\bibfnamefont{M.}~\bibnamefont{Hosoda}},
  \bibinfo{author}{\bibfnamefont{Y.}~\bibnamefont{Hikita}},
  \bibinfo{author}{\bibfnamefont{H.~Y.} \bibnamefont{Hwang}}, \bibnamefont{and}
  \bibinfo{author}{\bibfnamefont{K.~A.} \bibnamefont{Moler}},
  \bibinfo{journal}{Nat. Commun.} \textbf{\bibinfo{volume}{3}},
  \bibinfo{pages}{922} (\bibinfo{year}{2012}).

\bibitem[{\citenamefont{Lee et~al.}(2011{\natexlab{a}})\citenamefont{Lee,
  Clement, Hellerstedt, Kinney, Kinnischtzke, Leng, Snyder, and
  Goldman}}]{LeePhysRevLett2011PhasediagSTO}
\bibinfo{author}{\bibfnamefont{Y.}~\bibnamefont{Lee}},
  \bibinfo{author}{\bibfnamefont{C.}~\bibnamefont{Clement}},
  \bibinfo{author}{\bibfnamefont{J.}~\bibnamefont{Hellerstedt}},
  \bibinfo{author}{\bibfnamefont{J.}~\bibnamefont{Kinney}},
  \bibinfo{author}{\bibfnamefont{L.}~\bibnamefont{Kinnischtzke}},
  \bibinfo{author}{\bibfnamefont{X.}~\bibnamefont{Leng}},
  \bibinfo{author}{\bibfnamefont{S.~D.} \bibnamefont{Snyder}},
  \bibnamefont{and} \bibinfo{author}{\bibfnamefont{A.~M.}
  \bibnamefont{Goldman}}, \bibinfo{journal}{Physical Review Letters}
  \textbf{\bibinfo{volume}{106}}, \bibinfo{pages}{136809}
  (\bibinfo{year}{2011}{\natexlab{a}}).

\bibitem[{\citenamefont{Lee et~al.}(2011{\natexlab{b}})\citenamefont{Lee,
  Williams, Zhang, Frisbie, and Goldhaber-Gordon}}]{lee2011electrolyte}
\bibinfo{author}{\bibfnamefont{M.}~\bibnamefont{Lee}},
  \bibinfo{author}{\bibfnamefont{J.}~\bibnamefont{Williams}},
  \bibinfo{author}{\bibfnamefont{S.}~\bibnamefont{Zhang}},
  \bibinfo{author}{\bibfnamefont{C.}~\bibnamefont{Frisbie}}, \bibnamefont{and}
  \bibinfo{author}{\bibfnamefont{D.}~\bibnamefont{Goldhaber-Gordon}},
  \bibinfo{journal}{Physical Review Letters} \textbf{\bibinfo{volume}{107}},
  \bibinfo{pages}{256601} (\bibinfo{year}{2011}{\natexlab{b}}).

\bibitem[{\citenamefont{Moetakef et~al.}(2012)\citenamefont{Moetakef, Williams,
  Ouellette, Kajdos, Goldhaber-Gordon, Allen, and
  Stemmer}}]{moetakef2012carrier}
\bibinfo{author}{\bibfnamefont{P.}~\bibnamefont{Moetakef}},
  \bibinfo{author}{\bibfnamefont{J.~R.} \bibnamefont{Williams}},
  \bibinfo{author}{\bibfnamefont{D.~G.} \bibnamefont{Ouellette}},
  \bibinfo{author}{\bibfnamefont{A.}~\bibnamefont{Kajdos}},
  \bibinfo{author}{\bibfnamefont{D.}~\bibnamefont{Goldhaber-Gordon}},
  \bibinfo{author}{\bibfnamefont{S.~J.} \bibnamefont{Allen}}, \bibnamefont{and}
  \bibinfo{author}{\bibfnamefont{S.}~\bibnamefont{Stemmer}},
  \bibinfo{journal}{Physical Review X} \textbf{\bibinfo{volume}{2}},
  \bibinfo{pages}{021014} (\bibinfo{year}{2012}).

\bibitem[{\citenamefont{Sachs et~al.}(2010)\citenamefont{Sachs, Rakhmilevitch,
  Shalom, Shefler, Palevski, and Dagan}}]{Sachs2010S746}
\bibinfo{author}{\bibfnamefont{M.}~\bibnamefont{Sachs}},
  \bibinfo{author}{\bibfnamefont{D.}~\bibnamefont{Rakhmilevitch}},
  \bibinfo{author}{\bibfnamefont{M.~B.} \bibnamefont{Shalom}},
  \bibinfo{author}{\bibfnamefont{S.}~\bibnamefont{Shefler}},
  \bibinfo{author}{\bibfnamefont{A.}~\bibnamefont{Palevski}}, \bibnamefont{and}
  \bibinfo{author}{\bibfnamefont{Y.}~\bibnamefont{Dagan}},
  \bibinfo{journal}{Physica C: Superconductivity}
  \textbf{\bibinfo{volume}{470}}, \bibinfo{pages}{S746 }
  (\bibinfo{year}{2010}).

\bibitem[{\citenamefont{Ben~Shalom et~al.}(2009)\citenamefont{Ben~Shalom, Tai,
  Lereah, Sachs, Levy, Rakhmilevitch, Palevski, and
  Dagan}}]{shalom2009anisotropic}
\bibinfo{author}{\bibfnamefont{M.}~\bibnamefont{Ben~Shalom}},
  \bibinfo{author}{\bibfnamefont{C.}~\bibnamefont{Tai}},
  \bibinfo{author}{\bibfnamefont{Y.}~\bibnamefont{Lereah}},
  \bibinfo{author}{\bibfnamefont{M.}~\bibnamefont{Sachs}},
  \bibinfo{author}{\bibfnamefont{E.}~\bibnamefont{Levy}},
  \bibinfo{author}{\bibfnamefont{D.}~\bibnamefont{Rakhmilevitch}},
  \bibinfo{author}{\bibfnamefont{A.}~\bibnamefont{Palevski}}, \bibnamefont{and}
  \bibinfo{author}{\bibfnamefont{Y.}~\bibnamefont{Dagan}},
  \bibinfo{journal}{Physical Review B} \textbf{\bibinfo{volume}{80}},
  \bibinfo{pages}{140403} (\bibinfo{year}{2009}).

\bibitem[{\citenamefont{Ben~Shalom et~al.}(2010)\citenamefont{Ben~Shalom,
  Sachs, Rakhmilevitch, Palevski, and Dagan}}]{ben2010tuning}
\bibinfo{author}{\bibfnamefont{M.}~\bibnamefont{Ben~Shalom}},
  \bibinfo{author}{\bibfnamefont{M.}~\bibnamefont{Sachs}},
  \bibinfo{author}{\bibfnamefont{D.}~\bibnamefont{Rakhmilevitch}},
  \bibinfo{author}{\bibfnamefont{A.}~\bibnamefont{Palevski}}, \bibnamefont{and}
  \bibinfo{author}{\bibfnamefont{Y.}~\bibnamefont{Dagan}},
  \bibinfo{journal}{Physical Review Letters} \textbf{\bibinfo{volume}{104}},
  \bibinfo{pages}{126802} (\bibinfo{year}{2010}).

\bibitem[{\citenamefont{Caviglia et~al.}(2010)\citenamefont{Caviglia, Gabay,
  Gariglio, Reyren, Cancellieri, and Triscone}}]{caviglia2010tunable}
\bibinfo{author}{\bibfnamefont{A.~D.} \bibnamefont{Caviglia}},
  \bibinfo{author}{\bibfnamefont{M.}~\bibnamefont{Gabay}},
  \bibinfo{author}{\bibfnamefont{S.}~\bibnamefont{Gariglio}},
  \bibinfo{author}{\bibfnamefont{N.}~\bibnamefont{Reyren}},
  \bibinfo{author}{\bibfnamefont{C.}~\bibnamefont{Cancellieri}},
  \bibnamefont{and} \bibinfo{author}{\bibfnamefont{J.}~\bibnamefont{Triscone}},
  \bibinfo{journal}{Physical Review Letters} \textbf{\bibinfo{volume}{104}},
  \bibinfo{pages}{126803} (\bibinfo{year}{2010}).

\bibitem[{\citenamefont{Seri and Klein}(2009)}]{seri:180410}
\bibinfo{author}{\bibfnamefont{S.}~\bibnamefont{Seri}} \bibnamefont{and}
  \bibinfo{author}{\bibfnamefont{L.}~\bibnamefont{Klein}},
  \bibinfo{journal}{Phys. Rev. B} \textbf{\bibinfo{volume}{80}},
  \bibinfo{eid}{180410} (\bibinfo{year}{2009}).

\bibitem[{\citenamefont{Santander-Syro
  et~al.}(2011)\citenamefont{Santander-Syro, Copie, Kondo, Fortuna, Pailhes,
  Weht, Qiu, Bertran, Nicolaou, Taleb-Ibrahimi et~al.}}]{santander2011two}
\bibinfo{author}{\bibfnamefont{A.}~\bibnamefont{Santander-Syro}},
  \bibinfo{author}{\bibfnamefont{O.}~\bibnamefont{Copie}},
  \bibinfo{author}{\bibfnamefont{T.}~\bibnamefont{Kondo}},
  \bibinfo{author}{\bibfnamefont{F.}~\bibnamefont{Fortuna}},
  \bibinfo{author}{\bibfnamefont{S.}~\bibnamefont{Pailhes}},
  \bibinfo{author}{\bibfnamefont{R.}~\bibnamefont{Weht}},
  \bibinfo{author}{\bibfnamefont{X.}~\bibnamefont{Qiu}},
  \bibinfo{author}{\bibfnamefont{F.}~\bibnamefont{Bertran}},
  \bibinfo{author}{\bibfnamefont{A.}~\bibnamefont{Nicolaou}},
  \bibinfo{author}{\bibfnamefont{A.}~\bibnamefont{Taleb-Ibrahimi}},
  \bibnamefont{et~al.}, \bibinfo{journal}{Nature}
  \textbf{\bibinfo{volume}{469}}, \bibinfo{pages}{189} (\bibinfo{year}{2011}).

\bibitem[{\citenamefont{Meevasana et~al.}(2011)\citenamefont{Meevasana, King,
  He, Mo, Hashimoto, Tamai, Songsiriritthigul, Baumberger, and
  Shen}}]{meevasana2011creation}
\bibinfo{author}{\bibfnamefont{W.}~\bibnamefont{Meevasana}},
  \bibinfo{author}{\bibfnamefont{P.}~\bibnamefont{King}},
  \bibinfo{author}{\bibfnamefont{R.}~\bibnamefont{He}},
  \bibinfo{author}{\bibfnamefont{S.}~\bibnamefont{Mo}},
  \bibinfo{author}{\bibfnamefont{M.}~\bibnamefont{Hashimoto}},
  \bibinfo{author}{\bibfnamefont{A.}~\bibnamefont{Tamai}},
  \bibinfo{author}{\bibfnamefont{P.}~\bibnamefont{Songsiriritthigul}},
  \bibinfo{author}{\bibfnamefont{F.}~\bibnamefont{Baumberger}},
  \bibnamefont{and} \bibinfo{author}{\bibfnamefont{Z.~X.} \bibnamefont{Shen}},
  \bibinfo{journal}{Nature Materials} \textbf{\bibinfo{volume}{10}},
  \bibinfo{pages}{114} (\bibinfo{year}{2011}).

\bibitem[{\citenamefont{Xie et~al.}(2012)\citenamefont{Xie, Bell, Hikita, and
  Hwang}}]{xie2011tuning}
\bibinfo{author}{\bibfnamefont{Y.}~\bibnamefont{Xie}},
  \bibinfo{author}{\bibfnamefont{C.}~\bibnamefont{Bell}},
  \bibinfo{author}{\bibfnamefont{Y.}~\bibnamefont{Hikita}}, \bibnamefont{and}
  \bibinfo{author}{\bibfnamefont{H.~Y.} \bibnamefont{Hwang}},
  \bibinfo{journal}{Adv. Mater.} \textbf{\bibinfo{volume}{23}},
  \bibinfo{pages}{1744} (\bibinfo{year}{2012}).

\bibitem[{\citenamefont{Ngai et~al.}(2010)\citenamefont{Ngai, Segal, Su, Zhu,
  Walker, Ismail-Beigi, Le~Hur, and Ahn}}]{ngai2010electric}
\bibinfo{author}{\bibfnamefont{J.}~\bibnamefont{Ngai}},
  \bibinfo{author}{\bibfnamefont{Y.}~\bibnamefont{Segal}},
  \bibinfo{author}{\bibfnamefont{D.}~\bibnamefont{Su}},
  \bibinfo{author}{\bibfnamefont{Y.}~\bibnamefont{Zhu}},
  \bibinfo{author}{\bibfnamefont{F.}~\bibnamefont{Walker}},
  \bibinfo{author}{\bibfnamefont{S.}~\bibnamefont{Ismail-Beigi}},
  \bibinfo{author}{\bibfnamefont{K.}~\bibnamefont{Le~Hur}}, \bibnamefont{and}
  \bibinfo{author}{\bibfnamefont{C.}~\bibnamefont{Ahn}},
  \bibinfo{journal}{Physical Review B} \textbf{\bibinfo{volume}{81}},
  \bibinfo{pages}{241307} (\bibinfo{year}{2010}).

\bibitem[{\citenamefont{Biscaras et~al.}(2010)\citenamefont{Biscaras, Bergeal,
  Kushwaha, Wolf, Rastogi, Budhani, and Lesueur}}]{biscaras2010two}
\bibinfo{author}{\bibfnamefont{J.}~\bibnamefont{Biscaras}},
  \bibinfo{author}{\bibfnamefont{N.}~\bibnamefont{Bergeal}},
  \bibinfo{author}{\bibfnamefont{A.}~\bibnamefont{Kushwaha}},
  \bibinfo{author}{\bibfnamefont{T.}~\bibnamefont{Wolf}},
  \bibinfo{author}{\bibfnamefont{A.}~\bibnamefont{Rastogi}},
  \bibinfo{author}{\bibfnamefont{R.}~\bibnamefont{Budhani}}, \bibnamefont{and}
  \bibinfo{author}{\bibfnamefont{J.}~\bibnamefont{Lesueur}},
  \bibinfo{journal}{Nature Communications} \textbf{\bibinfo{volume}{1}},
  \bibinfo{pages}{89} (\bibinfo{year}{2010}).

\bibitem[{\citenamefont{Ye et~al.}(2009)\citenamefont{Ye, Inoue, Kobayashi,
  Kasahara, Yuan, Shimotani, and Iwasa}}]{ye2009liquid}
\bibinfo{author}{\bibfnamefont{J.}~\bibnamefont{Ye}},
  \bibinfo{author}{\bibfnamefont{S.}~\bibnamefont{Inoue}},
  \bibinfo{author}{\bibfnamefont{K.}~\bibnamefont{Kobayashi}},
  \bibinfo{author}{\bibfnamefont{Y.}~\bibnamefont{Kasahara}},
  \bibinfo{author}{\bibfnamefont{H.~T.} \bibnamefont{Yuan}},
  \bibinfo{author}{\bibfnamefont{H.}~\bibnamefont{Shimotani}},
  \bibnamefont{and} \bibinfo{author}{\bibfnamefont{Y.}~\bibnamefont{Iwasa}},
  \bibinfo{journal}{Nature materials} \textbf{\bibinfo{volume}{9}},
  \bibinfo{pages}{125} (\bibinfo{year}{2009}).

\bibitem[{\citenamefont{Kim et~al.}(2011)\citenamefont{Kim, Kozuka, Bell,
  Hikita, and Hwang}}]{kim2011intrinsic}
\bibinfo{author}{\bibfnamefont{M.}~\bibnamefont{Kim}},
  \bibinfo{author}{\bibfnamefont{Y.}~\bibnamefont{Kozuka}},
  \bibinfo{author}{\bibfnamefont{C.}~\bibnamefont{Bell}},
  \bibinfo{author}{\bibfnamefont{Y.}~\bibnamefont{Hikita}}, \bibnamefont{and}
  \bibinfo{author}{\bibfnamefont{H.~Y.} \bibnamefont{Hwang}},
  \bibinfo{journal}{Arxiv preprint arXiv:1106.5193}  (\bibinfo{year}{2011}).

\bibitem[{\citenamefont{Schneider et~al.}(2006)\citenamefont{Schneider, Thiel,
  Hammerl, Richter, and Mannhart}}]{schneiderlithography}
\bibinfo{author}{\bibfnamefont{C.~W.} \bibnamefont{Schneider}},
  \bibinfo{author}{\bibfnamefont{S.}~\bibnamefont{Thiel}},
  \bibinfo{author}{\bibfnamefont{G.}~\bibnamefont{Hammerl}},
  \bibinfo{author}{\bibfnamefont{C.}~\bibnamefont{Richter}}, \bibnamefont{and}
  \bibinfo{author}{\bibfnamefont{J.}~\bibnamefont{Mannhart}},
  \bibinfo{journal}{Applied Physics Letters} \textbf{\bibinfo{volume}{89}},
  \bibinfo{eid}{122101} (\bibinfo{year}{2006}).

\bibitem[{\citenamefont{Kozuka et~al.}(2009)\citenamefont{Kozuka, Kim, Bell,
  Kim, Hikita, and Hwang}}]{kozuka2009two}
\bibinfo{author}{\bibfnamefont{Y.}~\bibnamefont{Kozuka}},
  \bibinfo{author}{\bibfnamefont{M.}~\bibnamefont{Kim}},
  \bibinfo{author}{\bibfnamefont{C.}~\bibnamefont{Bell}},
  \bibinfo{author}{\bibfnamefont{B.}~\bibnamefont{Kim}},
  \bibinfo{author}{\bibfnamefont{Y.}~\bibnamefont{Hikita}}, \bibnamefont{and}
  \bibinfo{author}{\bibfnamefont{H.~Y.} \bibnamefont{Hwang}},
  \bibinfo{journal}{Nature} \textbf{\bibinfo{volume}{462}},
  \bibinfo{pages}{487} (\bibinfo{year}{2009}).

\bibitem[{\citenamefont{Bell et~al.}(2009)\citenamefont{Bell, Harashima,
  Kozuka, Kim, Kim, Hikita, and Hwang}}]{bell2009dominant}
\bibinfo{author}{\bibfnamefont{C.}~\bibnamefont{Bell}},
  \bibinfo{author}{\bibfnamefont{S.}~\bibnamefont{Harashima}},
  \bibinfo{author}{\bibfnamefont{Y.}~\bibnamefont{Kozuka}},
  \bibinfo{author}{\bibfnamefont{M.}~\bibnamefont{Kim}},
  \bibinfo{author}{\bibfnamefont{B.~G.} \bibnamefont{Kim}},
  \bibinfo{author}{\bibfnamefont{Y.}~\bibnamefont{Hikita}}, \bibnamefont{and}
  \bibinfo{author}{\bibfnamefont{H.~Y.} \bibnamefont{Hwang}},
  \bibinfo{journal}{Physical Review Letters} \textbf{\bibinfo{volume}{103}},
  \bibinfo{pages}{226802} (\bibinfo{year}{2009}).

\bibitem[{\citenamefont{Tang et~al.}(2003)\citenamefont{Tang, Kawakami,
  Awschalom, and Roukes}}]{tang2003giant}
\bibinfo{author}{\bibfnamefont{H.}~\bibnamefont{Tang}},
  \bibinfo{author}{\bibfnamefont{R.}~\bibnamefont{Kawakami}},
  \bibinfo{author}{\bibfnamefont{D.}~\bibnamefont{Awschalom}},
  \bibnamefont{and} \bibinfo{author}{\bibfnamefont{M.}~\bibnamefont{Roukes}},
  \bibinfo{journal}{Physical Review Letters} \textbf{\bibinfo{volume}{90}},
  \bibinfo{pages}{107201} (\bibinfo{year}{2003}).

\bibitem[{\citenamefont{V\'yborn\'y et~al.}(2009)\citenamefont{V\'yborn\'y,
  Ku\ifmmode~\check{c}\else \v{c}\fi{}era, Sinova, Rushforth, Gallagher, and
  Jungwirth}}]{Vybomy2009MicroscopicmechanismAMR}
\bibinfo{author}{\bibfnamefont{K.}~\bibnamefont{V\'yborn\'y}},
  \bibinfo{author}{\bibfnamefont{J.}~\bibnamefont{Ku\ifmmode~\check{c}\else
  \v{c}\fi{}era}}, \bibinfo{author}{\bibfnamefont{J.}~\bibnamefont{Sinova}},
  \bibinfo{author}{\bibfnamefont{A.~W.} \bibnamefont{Rushforth}},
  \bibinfo{author}{\bibfnamefont{B.~L.} \bibnamefont{Gallagher}},
  \bibnamefont{and}
  \bibinfo{author}{\bibfnamefont{T.}~\bibnamefont{Jungwirth}},
  \bibinfo{journal}{Phys. Rev. B} \textbf{\bibinfo{volume}{80}},
  \bibinfo{pages}{165204} (\bibinfo{year}{2009}).

\bibitem[{\citenamefont{Trushin et~al.}(2009)\citenamefont{Trushin,
  V\'yborn\'y, Moraczewski, Kovalev, Schliemann, and
  Jungwirth}}]{Trushin2009AMRpolaraizedspins}
\bibinfo{author}{\bibfnamefont{M.}~\bibnamefont{Trushin}},
  \bibinfo{author}{\bibfnamefont{K.}~\bibnamefont{V\'yborn\'y}},
  \bibinfo{author}{\bibfnamefont{P.}~\bibnamefont{Moraczewski}},
  \bibinfo{author}{\bibfnamefont{A.~A.} \bibnamefont{Kovalev}},
  \bibinfo{author}{\bibfnamefont{J.}~\bibnamefont{Schliemann}},
  \bibnamefont{and}
  \bibinfo{author}{\bibfnamefont{T.}~\bibnamefont{Jungwirth}},
  \bibinfo{journal}{Phys. Rev. B} \textbf{\bibinfo{volume}{80}},
  \bibinfo{pages}{134405} (\bibinfo{year}{2009}).

\bibitem[{\citenamefont{Joshua et~al.}(2011)\citenamefont{Joshua, Pecker,
  Ruhman, Altman, and Ilani}}]{joshua2011universal}
\bibinfo{author}{\bibfnamefont{A.}~\bibnamefont{Joshua}},
  \bibinfo{author}{\bibfnamefont{S.}~\bibnamefont{Pecker}},
  \bibinfo{author}{\bibfnamefont{J.}~\bibnamefont{Ruhman}},
  \bibinfo{author}{\bibfnamefont{E.}~\bibnamefont{Altman}}, \bibnamefont{and}
  \bibinfo{author}{\bibfnamefont{S.}~\bibnamefont{Ilani}},
  \bibinfo{journal}{Arxiv preprint arXiv:1110.2184}  (\bibinfo{year}{2011}).

\bibitem[{\citenamefont{Klemm et~al.}(1975)\citenamefont{Klemm, Luther, and
  Beasley}}]{SpinorbitHcpar}
\bibinfo{author}{\bibfnamefont{R.~A.} \bibnamefont{Klemm}},
  \bibinfo{author}{\bibfnamefont{A.}~\bibnamefont{Luther}}, \bibnamefont{and}
  \bibinfo{author}{\bibfnamefont{M.~R.} \bibnamefont{Beasley}},
  \bibinfo{journal}{Phys. Rev. B} \textbf{\bibinfo{volume}{12}},
  \bibinfo{pages}{877} (\bibinfo{year}{1975}).

\end{thebibliography}
\end{document}